\documentclass[letterpaper,twocolumn,10pt]{article}
\usepackage{usenix,epsfig,endnotes}
\usepackage{xcolor}
\usepackage{listings}
\usepackage{amsmath}
\usepackage{xfrac}

\newcommand{\mA}{\mathbf{A}} 

\definecolor{dkgreen}{rgb}{0,0.6,0}
\definecolor{gray}{rgb}{0.5,0.5,0.5}
\definecolor{mauve}{rgb}{0.58,0,0.82}

\lstdefinestyle{customc} {
    language=Python,numbers=left,keywordstyle=\bfseries\color{green!40!black},numberstyle=\color{gray} 
} 
 
\begin{document}

\date{}

\title{\Large \bf BuDDI: A Declarative Bloom Language for CALM Programming}

\author{
{\rm Rolando Garcia}\\
University of California, Berkeley\\
rogarcia@berkeley.edu
\and
{\rm Giulia Guidi}\\
University of California, Berkeley\\
Lawrence Berkeley National Laboratory\\
gguidi@berkeley.edu
}

\maketitle

\thispagestyle{empty}

\subsection*{Abstract}



Coordination protocols help programmers of distributed systems reason about the effects of transactions on the state of the system, but they're not cheap.
Coordination protocols may involve multiple rounds of communication, which can hurt system responsiveness.
There exist many efforts in distributed computing for managing the coordination-performance trade-off. 
More recent is a line of work that characterizes the class of workloads for which coordination is not necessary for consistency: namely, logically monotonic programs \cite{hellerstein2020keeping}. 
In this paper, we present a case study of logical monotonicity in workloads typical to computational biology. 
We leverage the Bloom language to write efficient distributed programs, and compare their performance to equivalent programs written in UPC++, a popular language for writing distributed programs.
Additionally, we leverage Bloom's analysis tools to identify points-of-coordination, and use our own experience using Bloom to recommend some higher-level abstractions for users without strong distributed computing backgrounds.

\vspace{-1em}
\section{Introduction}
\vspace{-.5em}

The rapid rise of cloud computing in recent years has brought exciting new challenges in the area of programming languages.
The overarching question is how we can make cloud programming easier, and thus make the cloud accessible to a wider range of applications.
The Bloom language is one of the leading efforts in this direction \cite{alvaro2011consistency}. 
Bloom is a language for disorderly distributed programming based on the principles of the CALM (consistency as logical monotonicity) theorem to guide coordination.
The CALM theorem states that coordination avoidance is possible when programs are  associativity, commutative, and idempotent.
The ability to implement an application according to these properties would naturally lead to a coordination-free implementation that would take greater advantage of cloud computing. 

We validate some of the assumptions of Bloom by asking the following questions: First, how common are logically monotonic programs in the wild? Specifically, in the area of computational biology. And second, does merely knowing where the coordination-points lie in a non-monotonic program help us rewrite the program to move coordination off the critical path or otherwise improve performance?
In this work, we borrow an application from computational biology (k-mer counting) and implement it in Bud (Bloom Under Development) and compare such an implementation with a standard UPC++ implementation.
In addition, we reflect on the design of the probabilistic data structure \emph{count min sketch} to reach a more space-efficient implementation.

Our work demonstrates that the assumptions of Bloom do in fact hold in practice, and Bloom is suitable for scientific computing.
Additionally, we leverage our experience using Bloom to propose some changes and a higher-level abstraction, one that is more use-friendly to programmers without strong distributed computing backgrounds.
In sum, the second core contribution of this paper is ``BuDDI'', an enhanced version of Bud using Distributed Data Independence (DDI).
BuDDI hides the challenges of distributed computing such as load balancing, data locality, and fault tolerance from the programmer so that even a novice programmer can take advantage of distributed computing.
This is critical as the explosion of data from many domains in recent years makes distributed computing a necessity rather than a commodity.
BuDDI uses the concept of global state and achieves the same goal of Bud in terms of coordination with comparable performance, while logically using global tables instead of local tables.

The paper is organized as follows. 
Section~\ref{sec:globaltab} describes the key global table concept in BuDDI and how it influences the design of BuDDI.
In Section~\ref{sec:iter} we describe another key construct, the iterator, while in Section~\ref{sec:null} we argue for a new design of the \texttt{Nil} concept for cloud programming.
Section~\ref{sec:kmerimpl} illustrates in detail the implementation of the k-mer counting case study as well as the design of the count min sketch, which is a probabilistic data structure  used for k-mer counting.
Finally, Section~\ref{sec:conclu} summarizes our conclusions and future work.

\vspace{-1em}
\section{Global Tables in BuDDI}\label{sec:globaltab}

In Bud, each worker can only read from and write to its own tables;
communication between workers is explicitly managed through channels.
By restricting tables to strictly local visibility,
Bud ensures that communication is only explicit when instructed by the developer, and is always asynchronous.
That is, a table update never triggers expensive consensus or communication protocols.
From our perspective, these design decisions make the Bud programmer responsible for both:
\begin{enumerate}

    \item The \textbf{efficient performance} of distributed computing: load balancing, straggler mitigation, data locality, resource disaggregation, auto-scaling, etc.
     \vspace{-.5em}
    \item The \textbf{intended behavior} of distributed computing: fault tolerance, fault detection, reliable communication, causal delivery, etc.
\end{enumerate}


By allowing the compiler and runtime to manage the data placement and communication of tuples stored in global tables, we significantly reduce the burden on the programmer.
By supporting global tables, we empower the compiler and runtime to manage caching, replication, sharding, and other data allocation decisions.
With a global table abstraction, the runtime environment can use online statistics and metadata to reorganize data as needed for latency, throughput, or reliability.
In this section we will show that it is possible to provide a global table abstraction without sacrificing the consistency and performance expected from Bloom languages.
In this work, we rely heavily on the work of CRDTs (conflict-free replicated data types) for inspiration ~\cite{shapiro2011conflict}.

\vspace{1em}
\subsection{Keep CALM and Merge Lattices}

In BuDDI, the global tables are CRDTs~\cite{shapiro2011conflict}.
This means that when a programmer inserts a tuple into a global table, the insertion succeeds immediately and the new tuple is shared asynchronously.
This ensures that BuDDI workers remain responsive even in the presence of network partitions.
The nature of CRDTs ensures that concurrent updates are eventually resolved.
A BuDDI programmer statically specifies the CRDT type of global tables, which can be:

\begin{itemize}
    \item\textbf{G-Set}: Grow-only set.
        \vspace{-.5em}
    \item \textbf{2P-Set}: Two-phase set comprised of positive and negative sets, each of which is a \texttt{G-Set}.
        \vspace{-.5em}
    \item \textbf{LWW-Set}: Last-writer-wins set represented as a timestamped \texttt{2P-Set}.
        \vspace{-.5em}
    \item \textbf{MV-Set}: Multi-value set, a dynamo-esque \texttt{2P-Set}.
\end{itemize}

By statically defining legal operations on global tables in advance, the developer enables the compiler and runtime to avoid unwanted coordination.
\texttt{G-Set}s are coordination free on non-monotonic queries,
and may wait to receive all tuples before computing a non-monotonic query.
\texttt{2P-Set}s may have to wait to receive all tuples in both the positive
and negative sets to service a monotonic read. The same 
coordination requirement applies when expressing equivalent computation in Bud, 
such as in the shopping cart example at checkout \cite{alvaro2011consistency}. 

The reader should note that tuples in global tables (i.e. CRDTs) are \textbf{not necessarily replicated by default}. 
The decision  whether to replicate (or cache or shard) depends on the programmer's service-level goals and is a customizable target for the compiler and runtime to hit.

\vspace{-0.75em}
\subsection{Compile-Time Coordination}

Whenever a Bud programmer organizes data such that the tuples are hash or range partitioned along the GROUP BY columns, the workers may compute aggregates without communication.
We refer to this communication-avoiding coordination strategy as  ``compile-time coordination'' because the programmer implicitly and statically informs every worker that data stored at other workers is irrelevant for the computation of the aggregate.
We will next show how the BuDDI compiler can manage data allocation to exploit compile-time coordination automatically in global tables. 


Explicit hash or range partitioning of data might be an easy task for Bud programmers, but fixed data partitioning logic prevents the runtime from adapting on the fly to changes in the number of workers, per-worker processing rate, and network conditions. In addition, the distribution of a data set is not always known in advance by the developer. This could lead to serious load balancing problems if the data is skewed.
Global tables allow the BuDDI compiler to parse the query to make data distribution decisions and inject switch operators into the IR so that the runtime can dynamically detect and adapt to changes.
For example, the data may be initially partitioned in a range, but after a skew is observed, the operators in the IR can switch to round-robin partitioning and inform the system of the need for communication-based coordination.
Such adjustments are not easy to express with only local tables and channels, but they follow naturally from the semantics of global tables based on CRDTs.

In Section~\ref{sec:Buddikmer}, we show a k-mer counting implementation in BuDDI that uses a global table of type \texttt{G-Set}, and relies solely on compile-time coordination.

\vspace{-.5em}
\subsection{One-Shot Fixed Point Computation}

Consider how a Bud programmer would remove items from a shopping cart:
\begin{equation}
    \small
    \texttt{
    shopping\_cart := shopping\_cart - bad\_items
    }
    \label{eq:1}
\end{equation}

Because \texttt{shopping\_cart} appears on the left hand side and right hand side of the assignment,
this is a recursive query that requires stratification: the query is evaluated repeatedly until
a fixed-point is reached \cite{green2013datalog}. 
It is possible to statically determine the need for stratification by detecting cycles in the dataflow graph.
As we will show in this section, it is not necessary to execute all recursive queries to a fix-point.
Some recursive queries, such as the one in Eq.~\ref{eq:1}, can be evaluated in one-shot.

It is possible to evaluate the recursive query in Eq.~\ref{eq:1} in one shot 
because the same query could be written using two \texttt{G-Set}:
\begin{equation}
    \small
    \texttt{
    shopping\_cart := added\_items - bad\_items 
    }
    \label{eq:2}
\end{equation}
The query is no longer recursive, and it is now statically possible
to determine that the query may be safely evaluated in one-shot, without the need for stratification.
In fact, this query is a manual implementation of a \texttt{2P-Set}, where wanted items are added to the 
``pos'' set and unwanted items are added to the ``neg'' set.
In BuDDI, the same query would be expressed using a \texttt{2P-Set}:
\begin{equation}
\small
\begin{split}
    & \texttt{shopping\_cart := 2P\_Set(`pos', `neg')} \\
    & \texttt{\dots} \\
    & \texttt{query(shopping\_cart)}
\end{split}
\end{equation}

If the semantics of a table provide the proper fit for a \texttt{2P-Set}, the programmer benefits from the following advantages when entering the table as a \texttt{2P-Set}:
\begin{enumerate}
    \item One-Shot Fixed-Point Computation: The runtime evaluates the contents of the table safely in one step.
    \item Async Garbage Collection: The runtime knows that it may safely delete elements from the pos-set iff it deletes the corresponding item from the neg-set.
    \item Compile-Time Coordination: The compiler knows that it \textbf{can} make data allocation decisions such that tuples in the same group are on the same worker in both the pos-set and the neg-set. In other words, the runtime could check the neg-set only at the local worker without communication and know if the tuple in the pos-set was deleted.
\end{enumerate}

\vspace{-0.75em}
\subsection{Zero-Knowledge Overwrite}\label{sec:zeroknow}

As our reader probably already knows, the neg-sets of \texttt{2P-Set} are tombstones.
Once a tuple is added to the neg- set, the same tuple can never be added to the \texttt{2P-Set} again.
Such a restriction violates the semantics of the set and makes it difficult or impossible to implement stateful applications.
A workaround for achieving stateful execution with CRDTs is to append a universally unique identifier (uuid) to each tuple.
This way, an object that has been removed from the \texttt{2P-Set} can be added back with a new uuid.
From now on we will refer to a \texttt{2P-Set} which models the set semantics as \texttt{True-Set}.
\texttt{True-Set} supports add, remove, add after remove, and update.

An advantage of the uuid approach is that it is possible to add tuples to a \texttt{True-Set} without coordination, since the probability of collisions guaranteed by cryptographically secure random number generators is negligible.
This quality we call \emph{zero-knowledge insertion}, because a worker may insert tuples into \texttt{True-Set} without knowing the content stored in replicas.

However, the uuid approach does not support zero-knowledge deletion or update. To update the contents of an object stored in a \texttt{ True-Set }, the worker must read the contents of the replicas to learn the uuid of the object. Blindly deleting the object could result in an anomaly where the new tuple is deleted because the messages may be reordered. 
Consequently, the delete operation must be parameterized by the uuid of the intended object and the request for the uuid must be served synchronously.

Using timestamps instead of uuid allows the support of updates and deletions with zero knowledge (in addition to insertions) for \texttt{True-Set}. 
The \texttt{True-Set} implements a variant of Last-Writer-Wins (LWW) or Multi-Value-Set (MV) to control concurrency.
For example, a \texttt{True-Set} implementing LWW would report the object with the latest timestamp.
Outdated versions of the object with old timestamps would be ignored or garbage collected.
A worker could delete an object from the \texttt{True-Set} without a synchronous read by inserting the object with a later timestamp into the tombstone.
If the timestamp of the object in the tombstone is greater than all the timestamps of the versions of that object in the pos-set, then the object is considered deleted.

Clock synchronization is not exactly a coordination-free affair, but such is the cost of sacrificing commutativity in CRDTs and imposing order in the cloud.


\vspace{-1em}
\section{Iterators}\label{sec:iter}

As global tables are to local tables, so are iterators to channels.
In this section, we will show that iterators are language constructs that can increase the declarativity of Bloom languages, especially with respect to stream or unstructured data processing.

\vspace{-.5em}
\subsection{Reading Big Files}

In early 2019, it was announced that a black hole had been photographed for the first time~\cite{akiyama2019first}.
Soon after, images of Dr. Katie Bouman, standing behind several stacks of hard drives, began circulating on Twitter.
The data needed to map the black hole weighed a total of $4.5$ petabytes. 
The question is, how can Bloom languages and the cloud in general even read such large files?

Database Management Systems can sort files of any size even with low memory specs. 
At the highest level, the key is to process the file incrementally, bringing it into memory one chunk at a time, and keeping track of the work that remains to be done.
Most of this behavior is provided by the iterator abstraction.
The caller invokes \texttt{next}, and the iterator returns a tuple or group of tuples (chunk), for processing.
The iterator ensures that the next chunk of work to be processed by the caller.


One approach for managing chunk size is to use as many chunks as there are workers---to enable high-throughput reads without requiring coordination.
Unfortunately, this approach is vulnerable to stragglers and cannot adapt on-the-fly to workers that enter or leave the workpool by default.
A workaround is to use a strategy similar to that used in MapReduce to allow the number of chunks to be much larger than the number of workers~\cite{dean2010mapreduce}.
Chunks are assigned to workers for delivery as they complete work.
Faster workers do more work than slower workers.
When new workers join, they can be assigned work at any time.
If a worker fails or leaves the group, data they have completed is not reassigned, and data they have not completed is returned to the pool of pending work.

\vspace{-.5em}
\subsection{Exactly-Once Semantics}

At least once delivery under set semantics 
is exactly-once semantics, because sets 
filter duplicates.
But as we discussed in Section~\ref{sec:zeroknow},
we have reason to give the same entity more than one uuid,
to enable re-insertion after deletion from a 2P-Set.

The solution then is to give each data unit (e.g. tuple)
two identifiers: one uuid identifies the token (e.g. the byte-offset from the start of the file);
the second uuid identifies the \emph{use} of the token.
Thus, an object that gets re-inserted to a 2P-Set would have the same token-uuid
as the one that is in the tombstone, but it would have a different \emph{use}-uuid.
We believe it's possible to statically assign token-uuids to units of data without coordination,
such as when we stamp each k-mer with its offset from the start of the file. Use-uuids are assigned
at runtime and may use either uuids or timestamps, as discussed in Section~\ref{sec:zeroknow},

With a token-uuid and use-uuid, it's possible to exploit at-least-once delivery
on CRDT True-Sets to achieve exactly-once semantics. All this with the benefits
of auto-scaling to workers entering or leaving the work group, or workers speeding up and slowing down.
\vspace{-.5em}
\section{Nil In The Cloud}\label{sec:null}

In developing BuDDI and implementing our case study in both BuDDI and Bud in the following section, the question arises as to how the programmer would interpret the vairable \texttt{Nil}.
In particular, the question is whether \texttt{Nil} tells us that the value does not exist for all queries and thus requires global coordination, or whether it tells us that the local worker is unable to find it.
The latter behavior is more appropriate for monotonic programming; although it might introduce anomalies, it can maintain responsiveness during a network partition.

One could argue that the \texttt{Nil} concept is an inherited appendage from the monolithic computer era. 
Here we argue that for cloud programming we should rethink the \texttt{Nil} concept as having not one but two values: either \texttt{DNE} (``does not exist'', which is a global assertion) or \texttt{IDK} (``I don't know'', which is a local assertion).
If the network is healthy and the data has been partitioned so that coordination is free or inexpensive (Section 2.2), the stronger property is considered and \texttt{DNE} is returned. 
Otherwise, \texttt{IDK} is returned.
We believe that \texttt{DNE} and \texttt{IDK} 
are sufficiently well understood by programmers to be useful boolean primitives.
For example, given a \texttt{DNE}, you could execute a block of code, whereas given a \texttt{IDK}, you would perform some other operation, such as retry, crash, sleep, or guess and apologize.

In Section~\ref{sec:kmerimpl} we will see how the \texttt{Nil} concept is not currently implemented in Bud and how its implementation would give the programmer more flexibility.
\vspace{-.5em}
\section{Case Study: K-mer Counter}\label{sec:kmerimpl}

A common computation in Computational Biology is to count the frequency of fixed length sequences known as k-mers.
The k-mer histogram that we obtain as a result is valuable for understanding the distribution of biological subsequences and for profiling genomic and metagenomic data.
For example, we may be interested in subsequences that occur within a certain interval or above a certain threshold.

The k-mer counting step often takes a large part of the total application runtime and it is a key computation within popular tools for taxonomic mapping~\cite{wood2014kraken}, metagenome classification~\cite{benoit2016multiple}, genome assembly~\cite{li2015megahit}.

The k-mer count is arguably a simple calculation, but its efficient implementation is anything but simple.
This problem has received much attention as an important target for shared memory parallelism.
As data sets grow faster and faster, distributed memory parallelization is becoming more and more important.
Nevertheless, the irregularity of the input data makes k-mer counting a difficult problem for distributed memory parallelization. In particular, the k-mer distribution over biological input data is not fixed and can only be determined at runtime.

\vspace{-.5em}
\subsection{Implementation}

In this work, we implement a toy version of the k-mer counting kernel in Bud and compare it to a UPC++ implementation, which resembles a more standard way of implementing this computation.
UPC++, similarly to Bud, makes use of asynchronous communication.
From a high level perspective, the main difference between the two codes is that the implementation based on Bud must follow Bloom's rules of monotony and idempotency.

\vspace{-.5em}
\paragraph{UPC++}

UPC++~\cite{bachan2017upc++} is a C++ library supporting Partitioned Global Address Space (PGAS) programming.
UPC++ is suited for implementing complex distributed data structures where communication is irregular or fine grained.
The main abstractions in UPC++ are: (a) global pointer to improve locality, (b) asynchronous remote procedure call (RPC), and (c) futures.

Listing~\ref{lst:upcxx} illustrates the key implementation for the UPC++ version of k-mer counter.
The k-mer counter materializes as a distributed hash table divided by keys, where k-mers are keys and their frequencies are values.
The program first parses the input data in parallel, so that each processor has a part of the input sequences.
Each processor parses its local sequences in k-mers and determines which k-mers remain local and which must be sent to another processor based on a hash function.
When a processor receives incoming data from other processors, it updates its local partition of the k-mer hash table by incrementing the frequency corresponding to the received k-mers.
A given k-mer is counted by one processor and only one processor.

Communication in UPC++ is asynchronous, and in our implementation we use a remote procedure call to update values in the hashmap that do not belong to the local processor.
A remote procedure call causes a procedure to be executed in a different address space, encoded like a normal procedure call, without the programmer explicitly coding the details for the remote interaction.

\begin{lstlisting}[language=C++, caption={UPC++ hashmap for k-mer counting.}, label={lst:upcxx}]
class DistrMap 
{
    /* <kmer, count> map */
    using dobj_map_t = dist_object<unordered_map<string, int>>;

    /* build empty map */
    dobj_map_t local_map{{}};

    /* compute owner for the given key */
    int get_target_rank(const string &key)
    {
        return hash<string>{}(key) % rank_n();
    };

    void local_update(unordered_map<string, int> &lmap, const string &key)
    {
        auto it = lmap.find(key);
        if(it != lmap.end()) it->second++; 
    };

    future<> populate(const string &key)
    {
        /* send rpc to the owner rank */
        return rpc(get_target_rank(key), 
        [](dobj_map_t &lmap, const string &key) 
        {
            /* check if key in local map */
            if(lmap->count(key) == 0) (*lmap)[key] = 1;
            /* update local value */
            else local_update(*lmap, key);
        }, local_map, key);
    };
};
\end{lstlisting}

\vspace{-.5em}
\paragraph{Bud}

Listings~\ref{lst:BudA}--\ref{lst:Buddyset} show different implementations that we developed for our case study in Bud.
In particular, the implementation in Listing~\ref{lst:BudB} uses domain-specific knowledge to optimize the memory footprint.

Counting subsequences is monotone by nature, because once you have seen a k-mer instance, the k-mer count can only increase or remain unchanged.
One might first think to use a \texttt{lmap} as a local partition, where k-mers are the keys and values are \texttt{lmax} lattices, since the k-mer count can only be incremented and \texttt{lmax} is defined as an integer that can only increment.
However, the default merge function of \texttt{lmax} is in fact the maximum between two entries, not the sum as we would wish.
It is not possible to override the merge function or create a custom lattice that uses sum as the merge function because sum is not idemponent.

The computation can be made idempotent by replacing \texttt{lmax} with \texttt{lset} in the local partition.
In this implementation, each k-mer \textit{instance} in the local partition is assigned a unique identifier. 
If ATAG occurs twice in the input data set, the corresponding \texttt{lset} in the local \texttt{lmap} stores two unique identifiers.
Once the computation is complete, we perform a local count of \texttt{lset} and display the k-mer frequency for each k-mer in the input data set.
In this work, we refer to this implementation as \texttt{Implementation A} (Listing~\ref{lst:BudA}).


\begin{lstlisting}[language=Ruby, caption={Bud k-mer counting: \texttt{Implementation A}. This implementation is not memory efficient because it stores as many uuid as k-mer instances.}, label={lst:BudA}]
# parse file and store sequences in an array
class DNA
  def readseq(myinput)
    sequences = Array.new
    ParseFasta::SeqFile.open(myinput).each_record do |rec|
      sequences.push(rec.seq.upcase)
    end
  end

  def kmers(sequences, k)

    subsequences = Hash.new
    for item in sequences do
        i = 0
        while i < item.length-k+1 do
            subsequences[SecureRandom.uuid] = item[i..k+i-1]
            i += 1
        end
    end

    karray = Array.new
    karray = subsequences.to_a
  end
end

class CountKmer
    include Bud
    state do
      scratch :kmer,      [:uuid, :seq]
      scratch :receive, [:seq, :uuid]
      scratch :leave,     [:seq, :uuid, :owner]
      lmap  :local
      lmap  :incoming
      lmap  :counter
      table :result
      channel :msg, [:seq, :uuid, :@addr]
    end

    bloom :ownership do
      # ip port based on sequence
      owner   = hash(t.seq) % worldsize
      leave <= kmer {|t| [t.seq, t.uuid, owner]}
      
      msg <~ leave do |seq, uuid, owner|
        [seq, uuid, owner]
      end
    end

    # update local set with incoming data
    bloom :insert do
      receive <= msg do |seq, uuid, owner|
        if owner == ip_port 
          [seq, uuid]
        end
      end

      # merge incoming kmer into local map
      incoming <= receive {|t| {t.seq => Bud::SetLattice.new([t.uuid])}}
      local     <= incoming
      counter   <= {ip_port => local}
    end

    bloom :count do
      result  <= counter.to_collection do |owner, m| 
          [owner, m.to_collection do |k, v| 
            [k, v.size]
          end
          ]
      end
    end
end
\end{lstlisting}

Implementation A is monotone and idempotent, however, it poses some concern about the memory consumption. 
In a medium-sized genome data set, we have billions of k-mers and some of these can occur hundreds of times in the input data set.
Therefore, storing an identifier for each k-mer instance looks extremely inefficient from a memory standpoint.
Fortunately, domain-specific knowledge helps us in the design of our \texttt{Implementation B} (Listing~\ref{lst:BudB}).
In general a user is only interested in subsequences that occur within a certain interval or below/above a certain threshold.
This information can be elegantly integrated into the Bud implementation using custom lattices.

\begin{lstlisting}[language=Ruby, caption={Bud k-mer counting: \texttt{Implementation B}. This implementation uses domain-specific knowledge to save memory and stores uuid only up to a certain threshold. Implementation B uses a custom lattice \texttt{Bud::BuddySetLattice} which is a modified version of the standard \texttt{Bud::SetLattice} reported in Listing~\ref{lst:Buddyset}.}, label={lst:BudB}]
class CountKmer
    include Bud
    [...]
    # update local set with incoming data
    bloom :insert do
      [..]
      # merge incoming kmer into local map
      incoming <= receive {|t| {t.seq => Bud::BuddySetLattice.new([t.uuid])}}
      local     <+ incoming
      counter   <+ {ip_port => local}
    end
    [...]
end
\end{lstlisting}

For example, if we are only interested in subsequences that occur above a certain threshold, we can create a custom \texttt{lset} where our merge function computes the union of two sets only if the reference set has not yet reached the custom threshold (Listing~\ref{lst:Buddyset}).
The threshold is commonly much smaller than the maximum frequency of high-frequency k-mers, making Implementation B much more memory efficient than Implementation A.

\begin{lstlisting}[language=Ruby, caption={\texttt{Bud::BuddySetLattice} is identical to the standard \texttt{lset} except for its \texttt{merge} function that uses domain-specific knowledge to avoid wasting memory.}, label={lst:Buddyset}]
THRESHOLD = 10
class Bud::BuddySetLattice < Bud::Lattice
  wrapper_name :buddylset
    [...]
    def merge(i)
      if @v.size < THRESHOLD
          wrap_unsafe(@v | i.reveal)
      else
          wrap_unsafe(@v)
      end
    end
    [...]
end
\end{lstlisting}

Before we approached the solution with a custom lattice, we tried an implementation with native lattices.
In Listing~\ref{lst:BudB-1} we use a naive \texttt{lset} lattice as the value in the \texttt{lmap} instead of the custom \texttt{Buddylset} in Listing~\ref{lst:Buddyset}.
In this case we want to add the received k-mer to the local \texttt{incoming} \texttt{lmap} only if we have not yet reached the threshold for this k-mer key.
The first error we encountered in implementing this version was a \textit{stratification} error, because when we merged \texttt{incoming} into \texttt{local} we used $\texttt{<=}$ instead of $\texttt{<+}$.
In this case, \texttt{incoming} merges with \texttt{local}, which in turn is used to calculate \texttt{incoming}.
To fix this error, we had to defer the merging of \texttt{incoming} into \texttt{local} with $\texttt{<+}$ until the next time step.
The stratification problem is solved at this point, but we have the problem that \texttt{local} may not contain the key we are looking up, and this caused a runtime error complaining that it cannot find the key.
Here, we wonder if it would be possible to return a \texttt{Nil} value when a key is missing, rather than causing a runtime error.
In Section~\ref{sec:null}, we briefly discussed how introducing two values (\texttt{DNE} and \texttt{IDK}) for the \texttt{Nil} concept would enable us to implement this implementation of k-mer counting without using a custom lattice. 
In particular, returning a local \texttt{IDK} would be sufficient in this case, since each k-mer (key) exists on one and only one process.

\begin{lstlisting}[language=Ruby, caption={Bud k-mer counting: \texttt{Implementation B} using only native types and related error.}, label={lst:BudB-1}]
THRESHOLD = 10
class CountKmer
    include Bud
    [...]
    # update local set with incoming data
    bloom :insert do
      [..]
      # merge incoming kmer into local map
      incoming <= receive {|t| {t.seq => Bud::SetLattice.new([t.uuid])} if (local.at(t.seq).size < THRESHOLD)} 
      local     <+ incoming
      counter   <+ {ip_port => local}
    end
    [...]
end
\end{lstlisting}

\vspace{-.5em}
\subsection{BuDDI}\label{sec:Buddikmer}

In this section, we demonstrate a hypothetical implementation 
of the k-mer Counting algorithm in BuDDI.
We use this example to demonstrate the use of G-Sets (a CRDT, or global table),
and iterators (e.g. the call to \texttt{open} the file).
The take-away from this subsection is that (i) BuDDI programs are simple for Bloom programmers to
write and understand, (ii) that their semantics are clear and unambiguous, and (ii) that the language is
sufficiently declarative to allow for acceptable performance with 
the aid of an intelligent compiler working in concert
with a specialized runtime. We leave both as future work.
For reference on achievable performance and semantics, 
the reader may want to refer back to Sections 2 and 3.

\begin{figure}[h]
    \centering
\begin{lstlisting}[language=Python, label={lst:k-merbuddy}, caption={BuDDI k-mer counting.}]
from buddy import Channel, GSet, View
from buddy import uuid, init

worker = Channel(key='@addr', value='file_uri')

kmers = GSet(key='uuid', value='seq')
result = View(query="""
                    SELECT seq, COUNT(*)
                    FROM kmers
                    GROUP BY seq
                    """)

def read_dna_file(kmer_stream = open(worker.file_uri, req_id=uuid(), addr=worker.addr, mode='char[4]')):
    global kmers
    kmers += {(uuid(), kmer) for kmer in kmer_stream}

@init
def register_worker(address: 'str', dna_file_uri: str):
    global worker
    worker += {(address, dna_file_uri)}
\end{lstlisting}
\end{figure}

The BuDDI programmer writes the k-mer counting program as
a SQL query with aggregation over a global table. 
In this case, the global table is called \texttt{kmers}, 
and it is implemented by a G-Set (or grow-only set).
Aggregation, when exact, is a non-monotonic operation,
but the runtime may begin processing the query on-the-fly,
and either report intermediate results or withhold them until the end as preferred by the user.

BuDDI executions allow for workers to join or leave the worker pool
on-the-fly. New workers are added with the \texttt{register\_worker} method
which monotonically inserts the worker's metadata into a channel.
BuDDI also uses an iterator (e.g. \texttt{open} in \texttt{read\_dna\_file})
to read the potentially massive file in parallel.
As we discussed in Section 3. 
The iterator is able to estimate network conditions and worker performance
by the rate at which workers request data from the iterator.
Thus, with the iterator, BuDDI is able to detect failures (a worker stops requesting data) and adapt on-the-fly 
to slowdowns. 

The BuDDI compiler statically recognizes that this workload is hash-partitionable on the \texttt{seq} column of the \texttt{kmers} table,
 and may rendezvouz tuples with matching token- uuids (no use-uuid required, since we are inserting into a G-Set),
into the same worker, so no communication is required to compute an aggregate. 
If the runtime does not replicate the data stored in the \texttt{kmers} CRDTs, then this execution and schedule achieves comparable performance to the hand-tuned Bud implementation, with all the additional benefits of declarativity.
\vspace{-.5em}
\subsection{Count Min Sketch Design}

Count min sketch~\cite{cormode2005improved} is a probabilistic data structure that serves as a frequency table of events in a stream of data. 
It is usually implemented as a data structure like a matrix, where $h$ rows represent your hash functions and $m$ columns represent the ranges, where $m$ is smaller than the number of k-mers because it is a sublinear data structure.
When inserting a k-mer $x$, this is hashed with the $h$ different hash functions and the counter in the corresponding column of the matrix calculated as hashed value module $m$ is incremented. 
And then we take the minimum count over the $h$ cells in the matrix where this k-mer was hashed.
In practice, $m$ is related to the actual number of k-mers, and usually people use the HyperLogLog algorithm~\cite{flajolet2007hyperloglog} to estimate the cardinality of the k-mer and chose $m$ accordingly.

The implementation of a count-min sketch is more complicated than that of a regular k-mer counting algorithm, since the data distribution is not straightforward and the data access pattern is not contiguous.
In this paper, we present two possible high-level designs and describe their current shortcomings.
The first design consists of the composition of standard Bloom lattices while the second design is based on a custom lattice that is more similar to the original data structure.

Using standard lattices, we can use a \texttt{lset} of size $m$, where each entry is a \texttt{lmap} of size $h$, where the key is the hash function id and the value is a \texttt{lset} of unique identifiers of k-mers; when merged, the \texttt{lset} increases its size by adding the corresponding identifier (uuid) to the entry [$m$]$\rightarrow$[$h$]$\rightarrow$[uuid].
In this case, the data distribution could be based on the $m$ ranges, where for $P$ processes each process has $\sfrac{m}{P}$ entries to take care of.
This distribution is relatively easy to implement, but the irregular data access pattern makes it more complicated to implement the \texttt{min} operator, because the $h$ entries on which we want to perform the operation could potentially belong to $h$ different processes.
This data structure and distribution requires cross-process coordination, and the communication pattern resembles an MPI \texttt{Reduce} or \texttt{Allreduce} collective communication.
The implementation of collective communication in BUD will remain as future work.
This first design trades coordination for memory usage, since each process has only one partition of the entire data structure, but may require coordination of all processes to reveal the final result.

A second design, based on a custom lattice, results in the opposite compromise: memory usage to reduce coordination.
The custom lattice is very similar to the original data structure, i.e. a $h \times m$ matrix $\mA$ (e.g. two Ruby Array structures), where each $\mA(i,j)$ entry is a k-mer unique identifier \texttt{lset}.
In this case, the distribution may be the same as we saw in the regular k-mer counting algorithm, i.e. each k-mer is hashed to a single process, and this process calculates the $h$ hash function on this k-mer and updates the corresponding matrix entries.
Here each process has a local copy of the entire matrix, and the merge operation is the \texttt{lset} merge operation applied to each of the matrix entries.
This means that consistency is ensured by an analysis at application level, which resolves write conflicts with a shared state.
This design allows easier distribution and less coordination, but the memory consumption is higher because each process has a local copy of the entire data structure.

The use of identical copies of the matrix on each process justifies the design and use of BuDDI and its logically global tables.
In BuDDI, the runtime may do replication or partitioning of data structures, but this is abstracted for the programmer.
In addition, BuDDI gives us access to the application-side code that we can use to learn the semantics of the application layer.
The application semantics may give BuDDI a way to resolve an apparent conflict, and can therefore allow BuDDI to reach good performance using weak consistency.

\vspace{-.5em}
\section{Conclusion and Future Work}\label{sec:conclu}

In this paper, we performed a case study of a representative HPC workload to evaluate the fitness of CALM programming for computational science.
In particular, we compared a Bud implementation with a UPC++ implementation and found that both are equally expressive.
Although a detailed performance comparison remains a future work, Bud provides enough low-level control to assume similar scaling behavior as UPC++.

However, Bud's low-level characteristics complicate distributed programming by offloading performance and correctness considerations to the user.
Our case study motivated the design of BuDDI, a \emph{more} declarative Bloom language that provides Distributed Data Independence. 
Our declarative abstractions rely on CRDTs and Iterators to maintain acceptable consistency and performance with the many benefits of declarative logic programming. 
Our hope is that the design of BuDDI will motivate our readers to work with us to implement the compiler and runtime.



\subsection*{Contributions}


RG worked primarily on the design of BuDDI, and discussion of Global Tables and Iterators.
GG worked primarily on the case study, implementation, and related discussion of k-mer counting in Bud and UPC++.
RG and GG contributed equally to the writing of the report.

This paper is submitted in fulfillment of the requirements of UC Berkeley's CS294 graduate seminar: ``Programming the Cloud.'' Taught by Professor Hellerstein and Dr. Milano.

\vspace{-.5em}
{\footnotesize \bibliographystyle{acm} \bibliography{usenix.bib}}

\begin{thebibliography}{10}

\bibitem{akiyama2019first}
{\sc Akiyama, K., Alberdi, A., Alef, W., Asada, K., Azulay, R., Baczko, A.-K.,
  Ball, D., Balokovi{\'c}, M., Barrett, J., Bintley, D., et~al.}
\newblock First m87 event horizon telescope results. iv. imaging the central
  supermassive black hole.
\newblock {\em The Astrophysical Journal Letters 875}, 1 (2019), L4.

\bibitem{alvaro2011consistency}
{\sc Alvaro, P., Conway, N., Hellerstein, J.~M., and Marczak, W.~R.}
\newblock Consistency analysis in bloom: a calm and collected approach.
\newblock In {\em CIDR\/} (2011), Citeseer, pp.~249--260.

\bibitem{bachan2017upc++}
{\sc Bachan, J., Bonachea, D., Hargrove, P.~H., Hofmeyr, S., Jacquelin, M.,
  Kamil, A., van Straalen, B., and Baden, S.~B.}
\newblock The upc++ pgas library for exascale computing.
\newblock In {\em Proceedings of the Second Annual PGAS Applications
  Workshop\/} (2017), pp.~1--4.

\bibitem{benoit2016multiple}
{\sc Benoit, G., Peterlongo, P., Mariadassou, M., Drezen, E., Schbath, S.,
  Lavenier, D., and Lemaitre, C.}
\newblock Multiple comparative metagenomics using multiset k-mer counting.
\newblock {\em PeerJ Computer Science 2\/} (2016), e94.

\bibitem{cormode2005improved}
{\sc Cormode, G., and Muthukrishnan, S.}
\newblock An improved data stream summary: the count-min sketch and its
  applications.
\newblock {\em Journal of Algorithms 55}, 1 (2005), 58--75.

\bibitem{dean2010mapreduce}
{\sc Dean, J., and Ghemawat, S.}
\newblock Mapreduce: a flexible data processing tool.
\newblock {\em Communications of the ACM 53}, 1 (2010), 72--77.

\bibitem{flajolet2007hyperloglog}
{\sc Flajolet, P., Fusy, {\'E}., Gandouet, O., and Meunier, F.}
\newblock Hyperloglog: the analysis of a near-optimal cardinality estimation
  algorithm.

\bibitem{green2013datalog}
{\sc Green, T.~J., Huang, S.~S., Loo, B.~T., Zhou, W., et~al.}
\newblock {\em Datalog and recursive query processing}.
\newblock Now Publishers, 2013.

\bibitem{hellerstein2020keeping}
{\sc Hellerstein, J.~M., and Alvaro, P.}
\newblock Keeping calm: when distributed consistency is easy.
\newblock {\em Communications of the ACM 63}, 9 (2020), 72--81.

\bibitem{li2015megahit}
{\sc Li, D., Liu, C.-M., Luo, R., Sadakane, K., and Lam, T.-W.}
\newblock Megahit: an ultra-fast single-node solution for large and complex
  metagenomics assembly via succinct de bruijn graph.
\newblock {\em Bioinformatics 31}, 10 (2015), 1674--1676.

\bibitem{shapiro2011conflict}
{\sc Shapiro, M., Pregui{\c{c}}a, N., Baquero, C., and Zawirski, M.}
\newblock Conflict-free replicated data types.
\newblock In {\em Symposium on Self-Stabilizing Systems\/} (2011), Springer,
  pp.~386--400.

\bibitem{wood2014kraken}
{\sc Wood, D.~E., and Salzberg, S.~L.}
\newblock Kraken: ultrafast metagenomic sequence classification using exact
  alignments.
\newblock {\em Genome biology 15}, 3 (2014), 1--12.

\end{thebibliography}


\end{document}